\documentclass[aps,prb]{revtex4}
\usepackage{graphicx}
\usepackage{bm}
\usepackage{xfrac}
\newcommand{\be}{\begin{equation}}
\newcommand{\ee}{\end{equation}}
\newcommand{\bea}{\begin{eqnarray}}
\newcommand{\eea}{\end{eqnarray}}
\newcommand{\beb}{\begin{eqnarray*}}
\newcommand{\eeb}{\end{eqnarray*}}

\begin{document}

\title{Exact wavefunctions for excitations of the $\nu =1/3$ fractional quantum Hall state from a model Hamiltonian}

\author{Paul Soul\'e}
\author{Thierry Jolicoeur}
\affiliation{Laboratoire de Physique Th\'eorique et Mod\`eles statistiques,
Universit\'e Paris-Sud, 91405 Orsay, France}

\date{November 10th, 2011}
\begin{abstract}
We study fractional quantum Hall states in the cylinder geometry with open boundaries.
We introduce solvable Hamiltonians for which we are able to obtain exact results. 
We give a simple construction of the ground state, quasiholes, quasielectrons and the magnetoroton branch of excited states
for spin-polarized electrons at filling factor $\nu=1/3$ and spinless bosons at filling $\nu=1/2$. 
The wavefunctions are simple in the second-quantized language.
These model Hamiltonians have all the features we expect from composite fermion theory.
\end{abstract}
\maketitle

\section{introduction}
Electrons in a two-dimensional gas state under a strong magnetic field display a wealth of
distinct states of matter. Among these are the incompressible liquids which manifest
the fractional quantum Hall effect (FQHE) with striking physical properties including the existence
of fractionally charged quasiparticles. The magnetic field perpendicular to the sample
leads to the formation of the macroscopically degenerate Landau levels whose filling
factor $\nu$ can be controlled by the electronic density and/or the magnetic field value.
The FQHE phenomenon appears for special rational values of $\nu$ where the electronic system forms 
an incompressible liquid with only gapped excitations. By changing the magnetic flux applied to
the system one may nucleate quasiholes or quasielectrons
Several theoretical approaches have to be combined to obtain an understanding of the FQHE.
Historically the physics of the most prominent state at $\nu=1/3$ 
in the lowest Landau level (LLL)
was understood by Laughlin
by writing down an explicit many-body wavefunction for the ground state~\cite{Laughlin83}. While this
Laughlin wavefunction is not the exact ground state of electrons interacting through
the Coulomb potential, it is nevertheless the ground state of a model Hamiltonian with
a hard-core potential and all evidence points to adiabatic continuity between this special
model and the physical situation. The physics can thus correctly be understood from studies
of the model wavefunction.
The wavefunction approach has been generalized successfully in the composite-fermion (CF)
construction~\cite{JainBook} for many of the observed fractions.
This constructions gives also trial wavefunctions for excited states.
However these are not eigenstates of any simple local Hamiltonian and they don't
become exact eigenstates in any known limit of the FQHE problem.
Even within the CF construction it is possible to use several slightly different
constructions that still capture the correct FQHE physics.
For the hard-core interaction, even if the Laughlin wavefunction is the exact ground state
then the simplest excited states, quasielectrons and magnetoroton states, are not known analytically.

From a practical point of view it is important to note that all these model wavefunctions
are formulated in the first-quantized language i.e. as explicit functions of particles coordinates
and that they are not simple in the natural language of standard many-body theory which is
the Fock space formulated through occupation numbers of quantum states. It implies
that computation of expectation values of observables has to be done by Monte-Carlo integration
rather than by analytical means.

While true samples displaying the electronic FQHE are small planar pieces of semiconductor devices,
it is theoretically useful to consider other geometries like the sphere or the torus. 
In this paper, we study the FQHE in the cylinder geometry which is compatible with the Landau gauge.
We impose periodic boundary conditions along the circumference $L$ of the cylinder and
the number of orbitals is finite due to a hard-wall condition at the ends of the cylinder. 
We consider both  spinless fermions and bosons with hard-core interactions in the LLL.
The physics is then governed by the ratio of the magnetic length $\ell=\sqrt{\hbar c/eB}$
and the finite size $L$ of the system. This quantity $\ell/L$ is analogous to the aspect ratio
of a torus. Indeed if we consider a FQHE droplet in the cylinder geometry it is squeezed into
a one-dimensional Luttinger-type system for $\ell/L\rightarrow 0$ while it is stretched into
a crystal-like state in the opposite limit $\ell/L\rightarrow \infty$. This was studied by
Rezayi and Haldane~\cite{RH94} by means of the Laughlin wavefunction for filling $\nu =1/3$.
The latter case is known as the Tao-Thouless~\cite{Tao83,Thouless84} or ``thin torus'' (TT) limit $L\rightarrow 0$.
At leading order the interactions become a problem of electrostatics~\cite{Chui86,BK,BHHK}.
The ground state is then a crystal state with a periodic pattern fixed by the filling factor $\nu$.
Many physical properties of this limit have to do with the FQHE physics but the relationship is
not complete~\cite{BK,BHHK}.
By truncating the hard-core interaction in powers of $\lambda=\exp (-2\pi^2\ell^2/L^2)$,
we construct simple Hubbard-like one-dimensional Hamiltonians  for which we are able to find infinitely many exact 
eigenstates. 
All these states are simply formulated in terms of occupation numbers i.e. in second-quantized language.
They are obtained by operating upon a ``root'' configuration with squeezing operators that bring closer
groups of particles. The root configuration is one of the electrostatic ground states found in the TT limit.
However the squeezing operation generates states that are no longer simple Slater determinants for
fermions or permanents for bosons.
Notably we give explicit formulas in second-quantization
for the quasielectron and the  magnetoroton on top of the fermionic $\nu=1/3$ ground state and its bosonic
analog at $\nu =1/2$. The root configuration of the magnetoroton shows that it can be viewed
as a quasihole-quasielectron bound state.
The quasiholes states are also exact eigenstates with the degeneracy given by their Abelian fractional statistics.
With a finite number of electrons in a finite area, the Fock space is finite-dimensional and 
It is then feasible to obtain numerically the spectrum of the Hamiltonian. This approach is valid
for arbitrary Hamiltonians including the Coulomb two-body interactions appropriate for
electrons in semiconductor
devices as well as hard-core interactions for neutral atoms in rapid rotation.
 We show by this method
that there is a range of values of the aspect ratio $\ell/L=O(1)$ for which the 
FQHE physics is recovered~: the CF scheme of many-body eigenstates is valid.
Interestingly the state counting is altered in the TT limit when looking at excited states~:
the TT limit is not smooth in general.

In section II, we explain the truncation scheme of the interactions. In section III, the construction of
exact eigenstates is detailed. In section IV, we give explicit second-quantized formulas for the quasielectron
and the magnetoroton. Section V discuss the relationship with FQHE physics by use of exact diagonalizations.
Finally section VI presents our conclusions.

\section{Truncation scheme}

We use the Landau gauge
with $A_x=0$ and $A_y=B x$ where $B$ is the strength of the applied magnetic field
and eigenstates can be taken with definite momentum along the $y$-axis. 
We consider spinless fermions or bosons in the lowest Landau level (LLL) and
impose periodic boundary conditions along the $y$ direction with a finite extent $L$~:
$\psi(y+L)\equiv \psi(y) $. The momentum $k$ is then quantized~: $k=2\pi n/L$
where $n$ is an integer. This defines the cylinder geometry, the radius of the cylinder being $L/2\pi$.
 The LLL one-body wavefunctions are given by~:
\begin{equation}
\phi_{n} (x,y)= 
\frac{Z^n \,\,{\lambda}^{{{ n^2}}}}{\sqrt{L {\ell} \sqrt{\pi} }}  \,\,
{\mathrm e}^{\displaystyle{- {x^2}/{2 {\ell}^2 }}},
\quad
 Z\equiv {\mathrm e}^{\frac{2\pi}{L}(x + i y) }\, .
\end{equation} 
It is important to note that the power $n$ of the complex $Z$ coordinates can be positive or negative.
A generic many-body wavefunction is thus a polynomial in the $Z$s \textit{and} $Z^{-1}$s of the particles~:
\be
\Psi (Z_1,\dots,Z_N)={\mathcal P}(Z_1,\dots,Z_N) 
\prod_i {\mathrm e}^{\displaystyle{- {x_i^2}/{2 {\ell}^2 }}} .
\ee
The Laughlin wavefunction in the cylinder geometry has been written by Rezayi and Haldane~\cite{RH94}~:
\be
\Psi^{(m)} =\prod_{i<j} \left(\frac{Z_i^\text{\sfrac{1}{2}}}{Z_j^\text{\sfrac{1}{2}}}
-\frac{Z_j^\text{\sfrac{1}{2}}}{Z_i^\text{\sfrac{1}{2}}}
\right)^m \, ,
\label{LPsi}
\ee
where we have omitted the ubiquitous exponential factor. The filling factor is then $\nu =1/m$
with $m$ odd (resp. even) for fermions (resp. bosons).

The Hilbert space is truncated by imposing $|n|\leq N_{max}$. Since
the Gaussian factor implies that there is spatial localization of orbitals the system has ``quasi'' hard walls
at  $|x|=2\pi N_{max}\ell^2/L$ and there are $2N_{max}+1$ orbitals. 
Incompressible FQHE states are realized for a special matching of the number of particles and the number of orbitals
which involves the so-called {shift} quantum number.
This set of boundary conditions breaks explicitly the translation symmetry.
It also creates two physical boundaries that can support edge modes of FQHE states.
With only one non-contractible loop around the cylinder, this geometry is different
from the previously well-studied sphere, torus or disk geometries.

We investigate the properties of hard-core interactions that are known to display the FQHE.
For spinless fermions this is the hard-core laplacian of delta
function introduced by Haldane~\cite{HaldaneBook}. 
There is ample numerical evidence that the Coulomb interaction
in the LLL shares the same FQHE physics as this hard-core limiting case. 
For spinless bosons this is the local delta function interaction which is an accurate representation 
of the s-wave scattering of ultracold bosonic atoms.

In second-quantized language we have in the fermionic case~:
\begin{equation}
 \mathcal H_F = \frac{1}{4}\sum_{\{n_i\}}  [(n_1-n_3)^2-(n_1-n_4)^2]\quad
\lambda^{ (n_1 - n_3)^2 + (n_1 - n_4)^2 }\quad
c^\dag_{n_1} c^\dag_{n_2} c_{n_3} c_{n_4} \, ,
\label{HamF}
\end{equation} 
while for bosons we find~:
\begin{equation}
 \mathcal H_B = \sum_{\{n_i\}}  \quad
\lambda^{ (n_1 - n_3)^2 + (n_1 - n_4)^2 }\quad
b^\dag_{n_1} b^\dag_{n_2} b_{n_3} b_{n_4} \, ,
\label{HamB}
\end{equation} 
where
the sum is restricted to $n_1+n_2=n_3+n_4$. Creation and annihilation operators are denoted by $ac_n, c_n^\dag$
for fermions and $b_n, b_n^\dag$ for bosons.
We have set the overall energy scale to unity.
Many-body eigenstates of this problem can be classified according to their total momentum
$K=\sum_i k_i$ along the $y$-direction. From now on, we measure the momentum in units of $2\pi/L$.
Note that the two-dimensional problem looks now like a one-dimensional chain of particles hopping
on sites indexed by the momentum $n$. This is due to the fact that the guiding center coordinates are
quantum-mechanically conjugate in the LLL.

The interaction can be naturally  expanded in powers of $\lambda$~:
\be
\mathcal H =\sum_n \lambda^n \mathcal H_n .
\ee
In the TT limit $\lambda\rightarrow 0$ we expect the physics to be dominated by the first few terms of this expansion~\cite{Chui86}.
The order $\lambda^{0,1}$ for bosons and $\lambda^{1,4}$ for fermions are electrostatic repulsion terms.
Such truncated Hamiltonians  without any hopping have been studied in detail~\cite{BK,BHHK}. They have crystalline ground states
whose structure is entirely defined by electrostatic energy considerations. They are a special case
of repulsive Hamiltonians obeying a convexity condition as a function of the range of the 
interactions~\cite{PU78,Hubbard78}. Such Hamiltonians display a devil staircase of ground states as a function of the filling
factor which is reminiscent of the FQHE series of fractions~\cite{BK,BHHK}. 
In the TT limit, the ground state is then a Slater determinant built from the classical minimum energy configuration (a permanent for bosonic states).
However the true FQHE problem, even for pure
hard-core interactions deviates from purely electrostatic models already at low order in $\lambda$ due to the appearance
of hopping with conserved center-of-mass~\cite{Seidel05}. It is thus natural to focus on  the truncated Hamiltonians that consistently
include the first nontrivial hopping terms. In the fermionic case we thus define~:
\begin{equation}
{\mathcal H}^{Fermi}_9=\lambda\sum_i n_i n_{i+1}
+4\lambda^4\sum_i n_i n_{i+2}
+9\lambda^9\sum_i n_i n_{i+3}
-3\lambda^5\left[\sum_i c_i^\dag c_{i+1}c_{i+2}c_{i+3}^\dag+h.c.\right].
\end{equation}
The corresponding Bose Hamiltonian is given by~:
\begin{equation}
{\mathcal H}^{Bose}_4=\sum_i n_i(n_i-1)
+4\lambda\sum_i n_i n_{i+1}
+4\lambda^4\sum_i n_i n_{i+2}
+2\lambda^2\left[\sum_i b_i^\dag b^2_{i+1}b_{i+2}^\dag+h.c.\right].
\end{equation}

In these equations the momenta are written as a site index $i$,
$b_i,b_i^\dag$ (resp. $c_i,c_i^\dag $) are bosonic (resp. fermionic) operators and $n_i$ is the occupation number. 
We have studied by extensive exact diagonalization these truncated models.
We find that there is a range of the aspect ratio for which the physics of CFs holds completely.
For example in fig.(\ref{F1}) we display the excited states at filling $\nu=1/3$, they are in excellent correspondence
with the full Hamiltonian. Some numerical evidence for CF/FQHE physics is discussed in section V.
Here we focus on exact eigenstates that can be obtained analytically in closed form.

\section{Exact eigenstates}
The Laughlin wavefunction Eq.(\ref{LPsi}) is the exact zero-energy ground state of the full
hard-core Hamiltonian Eqs.(\ref{HamF},\ref{HamB}). It is no longer an exact eigenstate
of the truncated Hamiltonians introduced above. However
the ground state at the same filling factor is still at zero-energy~\cite{Jansen}. The zero-energy property holds in fact
for all values of $\lambda$. It is readily explained by noting alternate formulas for the truncated models~:
\begin{equation}
{\mathcal H}^{Fermi}_9=\lambda\sum_i C_i^\dag C_i
+4\lambda^4\sum_i n_i n_{i+2}
+\lambda n_{\scriptscriptstyle{-N_{max}}} {n_{\scriptscriptstyle{1-N_{max}}}}+\lambda 
n_{\scriptscriptstyle{N_{max}}}n_{\scriptscriptstyle{N_{max}-1}},
\label{FT}
\end{equation}
\begin{equation}
{\mathcal H}^{Bose}_4=\sum_i B_i^\dag B_i
+4\lambda\sum_i n_i n_{i+1}
+n_{{\scriptscriptstyle{-N_{max}}}}(n_{\scriptscriptstyle{-N_{max}}}-1)
+n_{\scriptscriptstyle{N_{max}}} (n_{\scriptscriptstyle{N_{max}}}-1),
\label{BT}
\end{equation}
where we have defined~:
\be
C_i = c_{i+2}c_{i+1}+3\lambda^4 c_{i+3} c_{i},
\quad
B_i = b_{i+1}^2+2\lambda^2 b_i b_{i+2}.
\ee
It is the extended nature of the $B_i$ and $C_i$ operators that leads to the apparition of boundary terms
in Eqs.(\ref{FT},\ref{BT}). 

The ground state of the truncated Hamiltonian is given by~:
\be
\Psi_{GS} = { {\mathcal S}_F}|  1001001\dots 1001\rangle \, ,
\ee
in the Fermi case at $\nu=1/3$ and~:
\be
\Psi_{GS} = { {\mathcal S}_B}|  1010\dots 0101\rangle \, ,
\ee
in the Bose case at $\nu=1/2$
where ${\mathcal S}_{F,B}$ are \textit{squeezing} operators defined by~:
\begin{equation}
{\mathcal S}_F =\prod_n (1+3\lambda^4 c_{n-1}c_n^{\dag} c_{n+1}^\dag c_{n+2}),
\quad
{\mathcal S}_B =\prod_n (1-\lambda^2 b_{n-1}(b_n^{\dag})^2 b_{n+1}).
\end{equation}
It is a matter of trivial algebra to show that these states are annihilated by the ladder-like
operators $C_i$ and $B_i$ for all $i$ and then it follows immediately that they are zero-energy
eigenstates of  the truncated Hamiltonians Eqs.(\ref{FT},\ref{BT}) since the boundary terms
are also annihilated by the choice of the root configuration.
To accommodate a unique ground state root configuration in a finite number of orbitals
requires precisely the flux vs number of particles that includes a nontrivial shift~:
$2N_{max}=m(N-1)$ at filling $\nu=1/m$.
If we add more orbitals then we can construct a family of states by adding extra zeros at the boundaries of the system.
This is at no energy cost and is a reflection of the center of mass degeneracy in the LLL.

In the TT limit $\lambda\rightarrow0$ and the squeezings are suppressed. As a consequence,
 the state become simple Slater determinant (permanent for bosons).
For all other values of the aspect ratio these states are nontrivial polynomials that have the underlying
algebraic structure known for the Laughlin state~\cite{BH08}. 
These polynomials can be expanded in the monomial basis with the partial ordering known as ``dominance''
order on the particle configurations that are allowed.
In our case the squeezings are only nearest-neighbor
while the full Laughlin polynomial involves squeezing at arbitrary distances.
We note that more generally the Fock space decomposes into blocks through the action of the hopping term 
in Eqs.(\ref{HamF},\ref{HamB}) and that the stable subspaces
are much smaller than in the case of complete interactions, allowing exact diagonalization to reach larger system sizes.

If we add extra zeros anywhere in the root configurations, then these states remain zero-energy eigenstates.
They are the gapless quasihole excitations with the correct counting for the standard Laughlin state
with two edges. Indeed for $n$ quasiholes inserted into a root configuration the number of states is that
of $n$ bosons into $N$ orbitals which is the correct counting of Abelian quasiholes on the sphere.

It is simple to construct an infinite number of exact eigenstates of the truncated models
by considering~:
\be
\Psi = { {\mathcal S}}| {\mathrm{ root}} \rangle
\label{root}
\ee
where the root configuration is still annihilated by the ladder-like operators $C_i$, $B_i$.
For example it is possible to pile up bosons at both ends of the systems in arbitrary numbers.
Such a state will have high energy but will remain an exact eigenstate. In general we expect
that it will be the member of an excitation branch with extremal value of the total momentum.

\begin{figure}[htb]
\includegraphics[width=0.5\columnwidth]{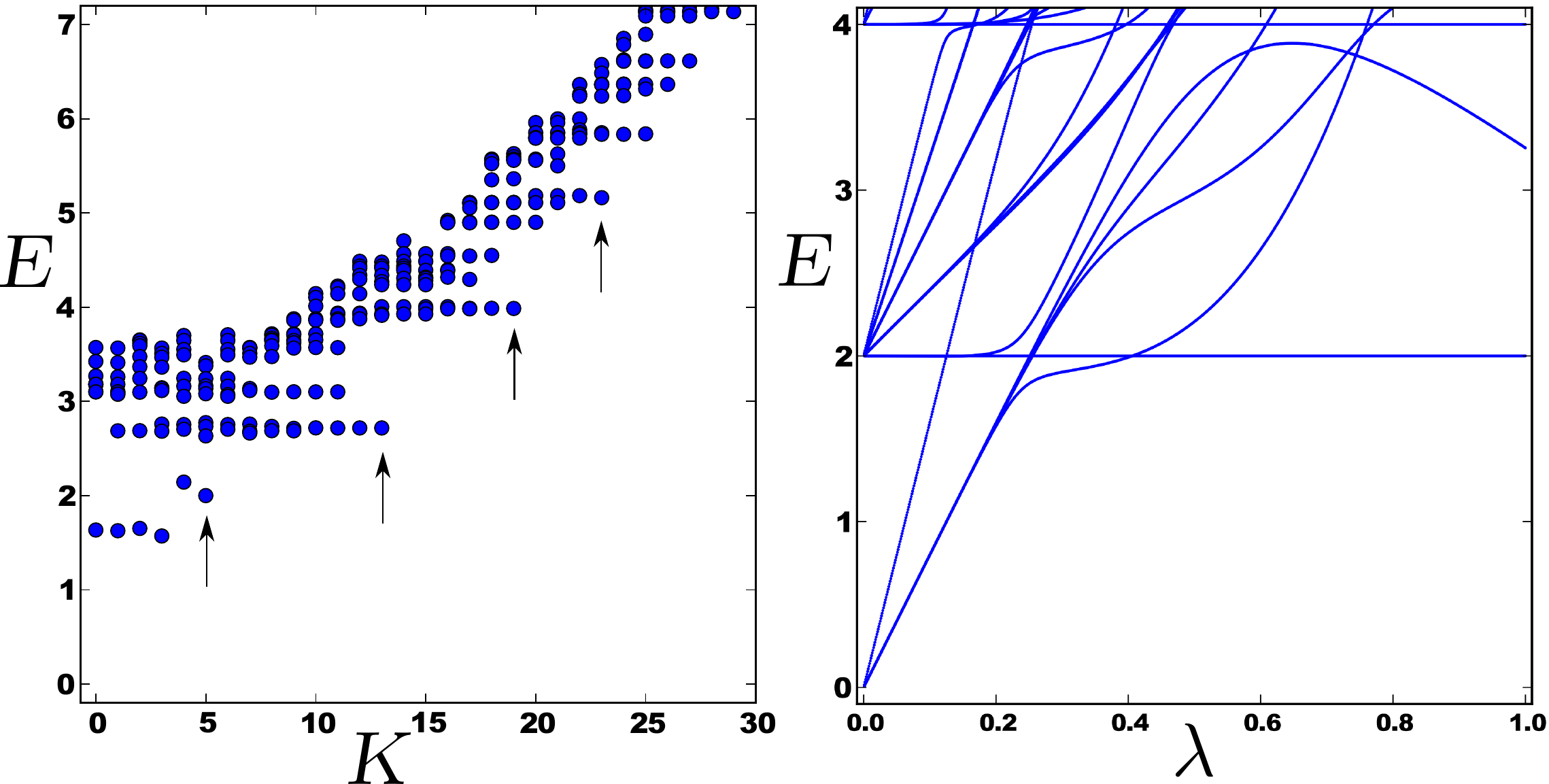}
\caption{The quasielectron appear as a set of almost degenerate levels spanning
$K=-N/2\dots +N/2$. Left panel is the spectrum of N=10 bosons with truncated interaction
and $\lambda=0.7$.
Arrows indicate states that are simple CF particle-hole excitations~\cite{JainBook}.
Right panel shows
the evolution of eigenstates at $K=N/2$ and N=5 (for clarity) as a function of $\lambda$.
The exact eigenstate Eq.(\ref{1QE}) has an energy independent of $\lambda$ and is no longer
the lowest energy state in the TT limit.}
\label{F2}
\end{figure}

\section{quasielectrons and magnetorotons}

The principal Laughlin fractions at $\nu =1/m$ fractions also support elementary 
gapped excitations in the form of the quasielectron which is obtained by
removing one flux quantum. On the sphere~\cite{Haldane} this state appears as a ground state with total angular momentum $L=N/2$.
On the cylinder we find accordingly a set of quasidegenerate states with $K_{tot}=-N/2\dots +N/2$
The member of this branch with extremal value of $K$ is given by~:
\begin{equation}
\Psi_{QE} = \mathcal{S}_F|11000100100\dots 001\rangle  \quad (\mathrm{Fermi}),
\label{1QE}
\end{equation}
\begin{equation}
 \Psi_{QE}=\mathcal{S}_B|2001010\dots 01\rangle\quad  (\mathrm{Bose}),
\end{equation}
and these states are readily seen to have energy $\lambda$ (resp. 2) in the Fermi (resp. Bose) case.
This is due solely to the boundary energies in Eqs.(\ref{FT},\ref{BT}).
This state cannot remain the lowest energy state in its momentum sector if we go to the TT limit because
end  points pay an energetic penalty. As a consequence, there are several level crossings when $L\rightarrow 0$~:
see Fig.(\ref{QEf}). More generally the CF counting rules that allow to identify multiplet of levels
are no longer valid in the TT limit. 

If we don't change the flux but stay at the special Laughlin filling $\nu =1/3$ the excited states
are dominated by the low-lying magnetoroton branch of excitations~\cite{GMP}. On the sphere it extends up
to $L=N$ and the extremal member is again an exact eigenstate given by~:
\begin{equation}
\Psi_{MR} = \mathcal{S}_F|11000100100\dots 0010\rangle \quad  (\mathrm{Fermi}),
\label{MR}
\end{equation}
\begin{equation}
 \Psi_{MR}=\mathcal{S}_B|2001010\dots 010\rangle\quad  (\mathrm{Bose}).
\end{equation}

This state has the same energy as the quasielectron since the only difference is that there is an added quasihole at the other extremity
of the system. This is exactly what we expect from the physical picture of the magnetoroton~\cite{GMP,KH84}.
This exciton branch is a quasielectron-quasihole bound state and the extremal member of the branch corresponds
to the maximal possible separation between the two elementary entities allowed by the geometry of the system.
It is possible to obtain a set of states with $K=N-1,N-2\dots$ that are degenerate and are members of the MR branch
by adding zeros in the root configuration above. These exact states do not fully exhaust the MR branch which is seen
to exist both in the hard-core model and in its truncation (see Fig.{\ref{F1}).
The other members of the MR branch are present in exact diagonalization studies but they are no longer very simple.
They are now inside a subspace spanned by the hopping term and this subspace is no longer of dimension one.

\begin{figure}
\includegraphics[width=0.5\columnwidth]{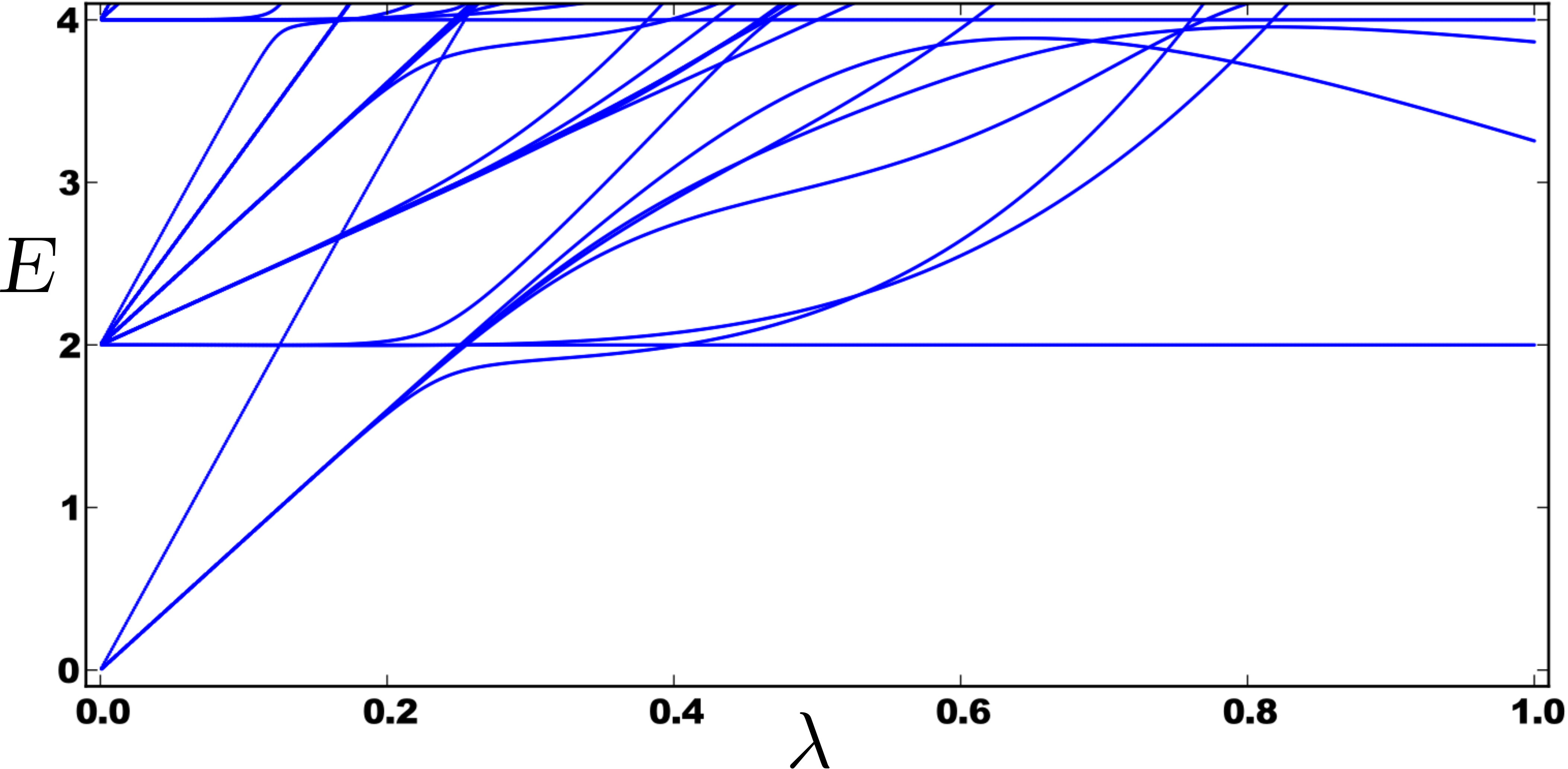}
\caption{Evolution of the low-lying levels as a function of $\lambda$ for $K=N=5$ bosons. The momentum is  chosen
to select the extremity of the magnetoroton branch. The exact wavefunction Eq.(\ref{MR})
is the ground state in this subspace isolated by a sizable gap from excited states.
for a large range of values of $\lambda$. However in the TT limit (left) there are many more low-lying states that no longer
obey the CF counting.}
\label{QEf}
\end{figure}

\section{composite fermion physics}

It remains to see if the special truncated Hamiltonians share the same FQHE as the complete
hard-core interactions. This is feasible by comparing the results of exact diagonalization studies for these
two cases. In the cylinder geometry the exact many-body eigenstates are classified by their total
momentum $K$ and the CF theory gives us simple rules for the appearance of specially stable states
in the full spectrum. This is the CF spectroscopy approach which has been successful in showing
that the Coulomb interactions shares the same FQHE physics as the hard-core interaction.

In fact we find that there is an exact mapping onto the many-body states
between the cylinder and the sphere given by $Z_i=u_i/v_i$ where
$u_i=\cos(\theta_i/2) {\mathrm e}^{i\phi_i/2},v_i=\sin(\theta_i/2) {\mathrm e}^{-i\phi_i/2}$ are the spinorial coordinates.
The two ends of the cylinder are mapped onto the two poles of the sphere. This may seem odd
because one may think that edge modes of the FQHE states live on the boundaries of the cylinder and hence
cannot be accommodated on the sphere.
However consideration of states with nonzero angular momentum leads to edge states~: the simplest example
being the state with \textit{maximal} momentum obtained by compressing electrons in a $\nu=1$ droplet at the North
pole. For generic flux this eigenstate has a boundary supporting the standard chiral boson modes.
If we use the relation from the sphere $2N_{max}=3(N-1)$ appropriate for the fermionic $\nu=1/3$ state then the boundary conditions on the cylinder
forbid low-energy excitations and we find an isolated ground state of zero energy~: see fig.(\ref{F1}). 

\begin{figure}
\includegraphics[width = 0.5\columnwidth]{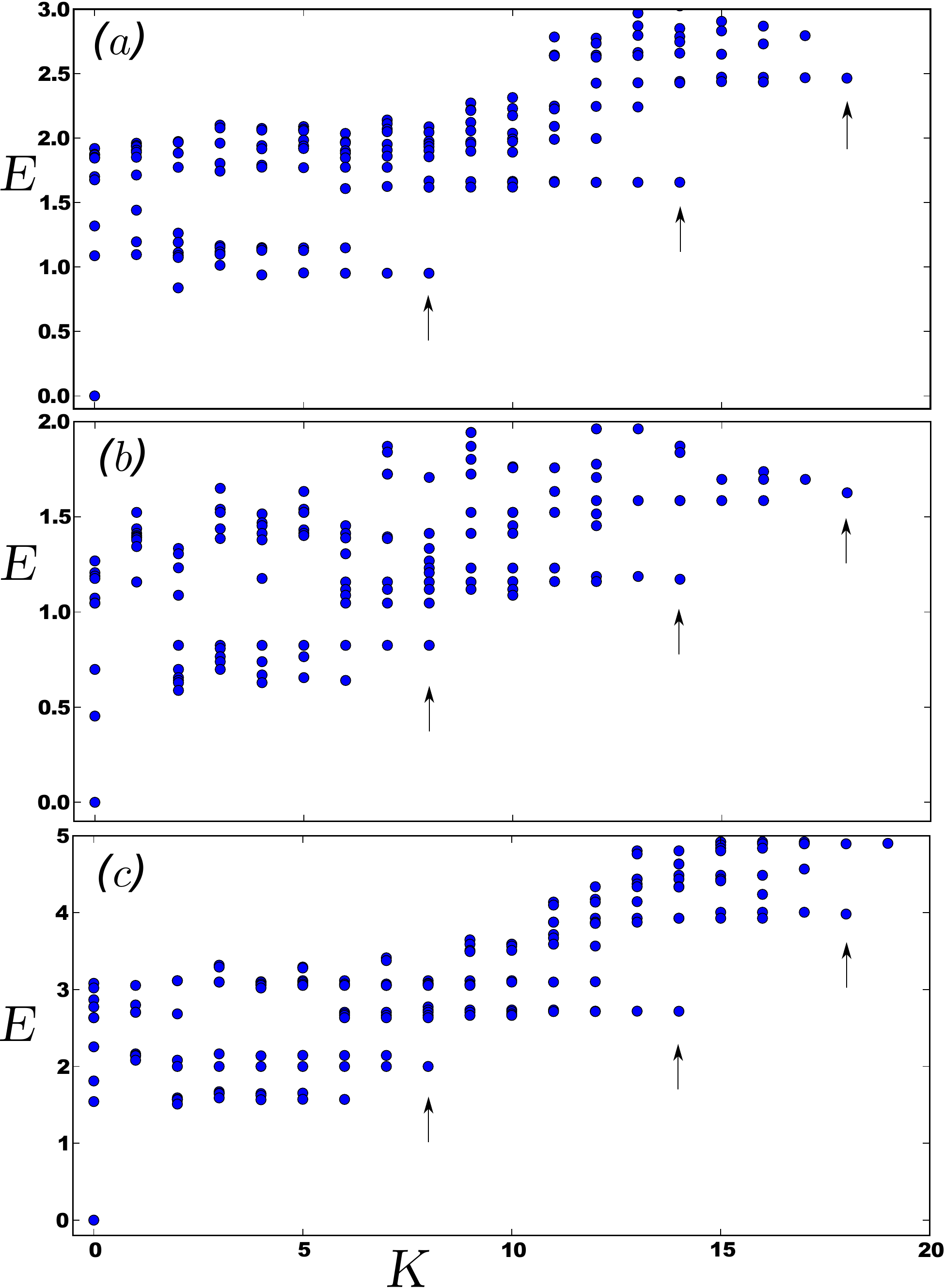}
\caption{The ten lowest-lying eigenstates as a function of the total momentum $K$ for N=8 particles. 
In (a) electrons are considered with the hard-core interactions at $\nu =1/3$ and $\lambda=0.82$.
The same problem with truncated interactions Eq.(\ref{FT}) is displayed in (b).
In (c) bosons with truncated interactions Eq.(\ref{BT}) at $\nu =1/2$ and $\lambda =0.7$.
Several branches of
composite fermion levels can be identified. Arrows show CF particle-hole states with extremal momentum~\cite{JainBook}. 
The left arrow points to the magnetoroton branch.}
\label{F1}
\end{figure}

We find generally that
there is a range of parameter $\lambda\sim 0.6-0.9$ where the CF scheme of levels is valid.
This interval is not universal, it is weakly dependent upon the number of particles as well as the Fermi/Bose
character of the particles involved.
We have constructed CF wavefunctions up to N=7 particles following Kamilla and Jain~\cite{KJ} and we obtain excellent agreement
between energies and overlap for the low-lying states. Quantitative details will be given in a forthcoming
publication.
In figure 1, we observe the appearance of the first branch of excitations corresponding to the promotion of one CF
into the second CF-Landau level extending up to $K_{max}=N$ the number of particles~:
this is the magnetoroton branch~\cite{GMP,KWJ} .
In the spherical geometry levels are classified by their total angular momentum. The exciton branch
has exactly one multiplet for $L_{tot}=N,N-1,N-2,\dots$. On the cylinder the conservation of $K_{tot}$
corresponds to the conservation of only the azimuthal angular momentum $L_z$ but the total momentum is broken by
the Hamiltonians (\ref{HamF},\ref{HamB}).

It is also possible to study other fractions of the hierarchy like $\nu=2/5$. We don't find simple exact states in this case.
While the gapped ground state readily appears in exact diagonalizations, the eigenstate is not simple. However the simple hopping
operator appearing in truncated problems leads to a reduced set of Fock states related by squeezing. The states created
by condensation of quasielectrons have the root configuration given by composite fermion wavefunctions~\cite{NBH09}~:
for the $\nu =2/3$ Bose state, we find that it is given by $|\mathrm{root}\rangle = |201011011\dots 0102\rangle $.
Again the TT limit is not smooth~: there are level crossings when $L\rightarrow 0$ even for the ground state
wavefunction. This is obvious since this root configuration contains double occupancy states that cannot go to low energy in the TT
limit (in the Fermi case it is two nearest-neighbor electrons at the end of the system)
This means that the cylinder geometry is different from this point of view from the torus geometry where there is adiabatic
continuity for ground states of many FQHE states~\cite{BK,BHHK}. Note that even in the torus states counting rules in the Haldane statistics
do change in the TT limit~\cite{KK}. A recent work~\cite{Nakamura} has proposed a construction related to ours albeit in a periodic chain geometry.

\section{Conclusion}

Starting from hard-core interactions between spinless fermions or bosons in the LLL
we have defined a truncated problem for which we have found
 infinitely many exact eigenstates for the Laughlin principal filling factors $\nu=1/m$.
These include some of the most important states of the FQHE physics~: the quasiholes, quasielectrons
and magnetorotons. These exact eigenstates are simple in second-quantized language
and can be manipulated easily for any system size analytically.
The quasihole-quasielectron bound state nature of the magnetoroton is manifest in this formulation.

To assess the relevance of our truncated problem to the real world,
we have numerically studied the FQHE on a cylinder with open boundaries and we have shown that it is closely related
to the spherical geometry. For a range of aspect ratio of the cylinder, we recover the FQHE physics 
as seen by comparison with the CF theory. This means that the simple exact states we have constructed capture
the main physical properties of the FQHE at these filling factors.

\begin{acknowledgments}
We acknowledge discussions with E. B. Bogomolnyi.
\end{acknowledgments}


\end{document}